\documentstyle[aps,prl,twocolumn]{revtex}
\def\be{\begin{equation}}
\def\ee{\end{equation}}
\def\bea{\begin{eqnarray}}
\def\eea{\end{eqnarray}}
\def\bma{\begin{mathletters}}
\def\ema{\end{mathletters}}
\newcommand{\bra}[1]{\mbox{$\langle #1 |$}}
\newcommand{\ket}[1]{\mbox{$| #1 \rangle$}}
\newcommand{\braket}[2]{\mbox{$\langle #1  | #2 \rangle$}}
\newcommand{\proj}[1]{\ket{#1}\!\bra{#1}}

\begin{document}

\draft

\title{Universality of optimal measurements.}

\author{Rolf Tarrach and Guifr\'e Vidal}

\address{e-mail: guifre@ecm.ub.es\\
Departament d'Estructura i Constituents de la
Mat\`eria\\ 
Universitat de Barcelona
\\ Diagonal 647, E-08028 Barcelona, Spain}

\date{\today}

\maketitle

\begin{abstract}
We present optimal and minimal measurements on identical copies of an unknown state of a qubit when the quality of measuring strategies is quantified with the gain of information (Kullback of probability distributions). We also show that the maximal gain of information occurs, among isotropic priors, when the state is known to be pure. Universality of optimal measurements follows from our results: using the fidelity or the gain of information, two different figures of merits, leads to exactly the same conclusions. We finally investigate the optimal capacity of $N$ copies of an unknown state as a quantum channel of information. 
\end{abstract}


Consider an unknown state of a two-level quantum system described by the density matrix $\rho(\vec{b})$, $\vec{b}$ being the Bloch vector, $b \equiv |\vec{b}| \leq 1$. The preparation device provides $N$ identical copies of the system, so that the state at our disposal is $\rho(\vec{b})^{\otimes N}$. In the past few years the optimal measuring strategy, i.e. the most successful at revealing the identity of the unknown state, has been obtained, first for pure states \cite{MasPop,DBE} and then for mixed states \cite{VLPT}. Also the minimal among the optimal strategies, i.e. the ones with the smallest number of outcomes, have been constructed, both for pure states \cite{LPT} and mixed states \cite{VLPT}. In the processing of information contained in quantum states, knowing the most efficient read-out procedures, i.e. the optimal and least resource consuming ones, is of course of importance.

 In all these contributions the quality of the measuring strategy, characterized by a resolution of the identity 
\be
\sum_i M_i = \openone,
\label{reso}
\ee
 in terms of positive operators $M_i \geq 0$, has been quantified by the fidelity \cite{Uhl}. In other words, when outcome $i$ (related to $M_i$) happens one guesses the unknown state to be $\tilde{\rho}_i \equiv \rho(\vec{p}_i)$ and one quantifies the quality of the guess by

\be
F(\rho(\vec{b}),\rho(\vec{p}_i)) \equiv \left(\mbox{Tr}[~\sqrt{\rho(\vec{b})^{1/2}\rho(\vec{p}_i)\;\rho(\vec{b})^{1/2}}~]~\right)^2.
\label{fid}
\ee

 One can arrive at eq. (\ref{fid}) from several different starting points. One of them is based on a measure of distinguishability of the probability distributions associated to $\rho$ and $\rho'$ by performing general positive operator valued measurements (as in eq. (\ref{reso})) on them \cite{Woo} and minimizing,

\be
F(\rho, \rho') = \min 
\left( \sum_j \sqrt{\mbox{Tr}[\rho M_j]}\sqrt{\mbox{Tr}[\rho' M_j]}\right)^2.
\ee

Another is based on the standard Hilbert space scalar product of the two pure states which belonging to ${\cal C}^2\otimes{\cal C}^2$ lead to $\rho$ and $\rho'$ when reduced \cite{Jos},

\be
F(\rho, \rho') = \max |\braket{\psi}{\psi'}|^2, 
\ee
where maximization is performed over $\{\ket{\psi}, \ket{\psi'}\} / \rho = \mbox{Tr}_a [\proj{\psi}],~\rho' = \mbox{Tr}_a [\proj{\psi'}].$

 These equivalent definitions of the fidelity, plus the following properties which characterize it further, make it a unique quantification of the comparison of two general quantum states:
\begin{enumerate}
\item $0 \leq F(\rho,\rho') = F(\rho',\rho) \leq 1$.
\item $F(\rho,\rho')= 1 ~\Leftrightarrow~ \rho = \rho';~~ F(\rho,\rho')= 0 ~\Leftrightarrow~ \rho\rho'=0$.
\item $F(U\rho~ U^{\dagger},U\rho'U^{\dagger})=F(\rho,\rho'), ~~UU^{\dagger}=U^{\dagger}U=\openone$.
\item $F(\proj{\psi},\rho) = \bra{\psi} \rho \ket{\psi}$.
\item $F(\rho\otimes \sigma,\rho'\otimes\sigma') = F(\rho,\rho')F(\sigma,\sigma')$
\item $F(\rho,p\rho_1\!+\! (1\!-\!p)\rho_2) \geq pF(\rho,\rho_1) + (1\!-\!p) F(\rho,\rho_2),\\ 0\leq p\leq 1$. 
\end{enumerate}

 In references \cite{MasPop,DBE,LPT} the unknown state was known to be pure, $b=1$, but no knowledge of the direction of the Bloch vector was assumed. In reference \cite{VLPT} the unknown state was a mixed state drawn stochastically from a known isotropic distribution $f(b)$, and although the best guess $\tilde{\rho}_i$ depended on $f(b)$, the optimal measuring strategy, that is the set $\{ M_i \}$ of positive operators of the different outcomes, did not. For isotropic distributions optimal measurements are thus distribution, i.e. $f(b)$, independent.

 However, proposing an outcome-dependent guess and evaluating its quality through the fidelity is only one of the criteria that could have been used to define optimal measurements. A sound alternative, the one we shall investigate in this work and probably the most sensible choice in the context of quantum information theory, consists of quantifying the quality of measuring strategies through the gain of information about the unknown state. In fact, information theory already supplies a universally accepted, unambiguous scheme for this purpose, that we shall follow. It is based on Bayes formula, which provides a conditional (outcome-dependent), posterior distribution $f_c(\vec{b}|i)$ from the (here isotropic) prior distribution $f(b)$, and on the Kullback, which quantifies the gain of information acquired when replacing $f(b)$ with $f_c(\vec{b}|i)$.

 More specifically, if $P_i(\vec{b}) \equiv \mbox{Tr}[\rho(\vec{b})M_i]$ is the probability of outcome $i$ when the unknown state is $\rho(\vec{b})$ and 
\be
P_{ap}(i) \equiv \int d^3b~f(b) P_i(\vec{b})~~~(\int d^3b~f(b) = 1)
\label{apriori}
\ee
is the a priori probability of outcome $i$, then Bayes formula states that the posterior distribution $f_c(\vec{b}|i)$, the one which collects our knowledge about the unknown state $\rho(\vec{b})$ after measuring when the initial knowledge was given by $f(b)$, reads
\be
f_c(\vec{b}|i) = \frac{f(b)P_i(\vec{b})}{P_{ap}(i)}.
\label{Bayes}
\ee 
 The gain of information about $\rho(\vec{b})$, $\Delta I$, is then given, in bits, by the Kullback of $f_c(\vec{b}|i)$ relative to $f(b)$ \cite{Kull}
\be
K_i[f_c/f] \equiv  \int d^3b~ f_c(\vec{b}|i) \log_2 \frac{f_c(\vec{b}|i)}{f(b)}.
\label{kullback}
\ee
This expression, the only one satisfying a series of intuitively reasonable conditions \cite{Hob}, is well-defined for continuous distributions (it has no dependence on the measure in the space of quantum states) and its average over possible outcomes,
\be
\bar{K}[f_c/f] \equiv \sum_i P_{ap}(i) K_i[f_c/f],
\label{avekullback}
\ee
is precisely the difference of the a priori and average a posteriori entropies  $H$ of the corresponding probability distributions of states,
\bea
H[f]-\bar{H}[f_c] \equiv - &\int& d^3b~ f(b) \log_2 f(b) \nonumber \\
 + \sum_i P_{ap}(i) &\int& d^3b~ f_c(\vec{b}|i)\log_2 f_c(\vec{b}|i),
\label{deltaentro}
\eea
as can be checked by considering eqs. (\ref{Bayes}-\ref{avekullback}) and that $\sum_i P_i(\vec{b})=1$ \cite{text}. This quantification is therefore equivalent to the one already used in previous works on quantum state estimation with discrete distributions (see, e.g., ref. \cite{PerWoo}).

 First, the question of which are the optimal measurements according to this information theoretically based criterion will be addressed. We will check explicitly for $N=1$ and $N=2$, and provide clues for any $N$, that optimal --and also minimal-- measuring strategies are universal, i.e. independent of whether the fidelity or the increase of information is used for their quantification, and will compute the corresponding optimal gain of information $\Delta I$. Then we will move to consider which is the isotropic prior $f(b)$ for which optimal measurements extract most information, so that it corresponds to the optimal (isotropic) quantum channel of information. After introducing a reversible compression procedure we conclude that the optimal amount of extractable information is, as $N\rightarrow\infty$, of one bit per effective qubit isotropic distributions.

In order to find an optimal measuring strategy, i.e. a set of operators $M_i$ as in eq. (\ref{reso}) maximizing the gain of information (eq. (\ref{avekullback})), the following theorem and subsequent corollaries, valid for any number of copies $N$, will be very useful.

\vspace{2mm}

\noindent{\bf Theorem:} Let the positive operator $M_i \geq 0$ be such that its probability $P_i(\vec{b})= \mbox{Tr}[M_i\rho(\vec{b})^{\otimes N}]$ can be written, for any $\vec{b}$, as the sum of two contributions of the form 
\[
P_{i,k}(\vec{b})\equiv$Tr$[M_{i,k}\rho(\vec{b})^{\otimes N}], ~~k=1,2,
\]
where the operators $M_{i,1},M_{i,2}$ are also positive (and $M_{i,1}+M_{i,2}$ is not necessarily equal to $M_i$). Let us introduce corresponding prior probabilities $P_{ap}(i,k)$ and posterior distributions $f_c(\vec{b}|i,k)$ as in eqs. (\ref{apriori}) and (\ref{Bayes}). Then, 
\be
P_{ap}(i) K_i[f_c/f] \leq \sum_{k=1}^2 P_{ap}(i,k) K_{i,k}[f_c/f].
\ee
\vspace{3mm}

\noindent{\bf Proof:} It follows from the inequality
\be
(x_1+x_2)\ln\frac{x_1+x_2}{y_1+y_2} \leq x_1\ln \frac{x_1}{y_1} +  x_2\ln \frac{x_2}{y_2}
\ee
$\forall~ x_1,x_2,y_1,y_2 \geq 0$. $\Box$
\vspace{3mm}

\noindent{\bf Corollary 1:} An optimal measuring strategy with rank-one operators always exists. (cf.\cite{Dav})
\vspace{3mm}

\noindent{\bf Proof:} Indeed, suppose $\sum_i M_i = \openone$ corresponds to an optimal measurement. Then, if $M_i = \sum_{k} \proj{i,k}$ is the spectral decomposition of $M_i$, it follows from the theorem that the rank-one POV measurement $\sum_{i,k} \proj{i,k} = \openone$ is also optimal. $\Box$
\vspace{3mm}

 We can already consider the case $N=1$, that is, when only one copy of the unknown state is available. One can convince oneself immediately that an optimal (and also minimal) measurement is just a standard von Neumann measurement. In fact, any will do because of the isotropy of $f(b)$. Suppose that we measure $\sigma_z$. Then, for $\vec{b}= (b\sin\theta\cos\phi,b\sin\theta\sin\phi, b\cos\theta)$, we have
\be
f_c(\vec{b}|\pm) = (1\pm b\cos \theta ) f(b)
\ee
and the gain of information is 
\bea
\Delta I^{(1)}=\pi\int_0^1 db~b^2 f(b)\left[\frac{(1+b)^2}{b}\log_2(1+b)-\right. \nonumber\\
 \left.\frac{(1-b)^2}{b}\log_2(1-b)\right] ~~-~~\frac{\log_2 e}{2}.~~~~~~~~~
\label{info1}
\eea

The function in square brackets in eq. (\ref{info1}) is monotonically increasing, so that the distribution for which the absolute increase in knowledge is maximal is
\be
f_m^{(1)}(b) = \frac{1}{4\pi}\delta(b-1),
\label{pure}
\ee
i.e. an isotropic distribution of pure states.

 It is interesting to point out that if instead of using in ref. \cite{VLPT} the mean average fidelity $\bar{F}^{(1)}$ we had used the mean average increase in fidelity,
\be
\Delta F^{(1)} \equiv \bar{F}^{(1)} -  F^{(1)}_{ap},
\ee
with the optimal guess $\tilde{\rho}_0 \equiv \rho(0)$ if no measurement is performed, so that
\be
 F^{(1)}_{ap} = \frac{1}{2} + I_{1/2} = F_{ap}^{(N)}
\ee
with (cf. \cite{VLPT})
\be
I_{\alpha} \equiv 4\pi \int_0^1 db~b^2 f(b)(\frac{1-b^2}{4})^{\alpha}
\ee
($I_0 = 1, I_{\alpha} \geq 4I_{\alpha+1}$), we would have obtained 
\be
\Delta F^{(1)} = \sqrt{I_{1/2}^2+\frac{1}{36}(1-4I_1)^2} - I_{1/2}.
\ee
It is then easily verified that the maximum value of $\Delta F^{(1)}$ also corresponds to the distribution eq. (\ref{pure}). Thus, for N=1, quantifying with the fidelity or with the Kullback information leads to the same (for N=1 somewhat obvious) optimal and minimal measuring strategy and to the same distribution which maximizes $\Delta I^{(1)}$ and $\Delta F^{(1)}$. Is this also true for $N=2$?

 In order to answer this question we need to present a second corollary. Notice first that with the following notation (borrowed from \cite{VLPT}) for the composite Hilbert
space of $N$ copies of the unknown state $\rho(\vec{b})$,
\begin{equation}
\label{hilberts}
{\cal H}^{(N)} \equiv {\cal H}_A \otimes
{\cal  H}_B \otimes ... {\cal H}_N,
\end{equation}
for the corresponding local spin operators,
\begin{eqnarray}
\label{sabn}
\nonumber
\vec S_A &\equiv& {1\over 2}\vec {\sigma} \otimes I^{\otimes N-1},\\
\nonumber
\vec S_B &\equiv& {1\over 2} I \otimes \vec {\sigma} \otimes
I^{\otimes N-2},\\
\vdots \nonumber\\
\vec S_N
 &\equiv& {1\over 2} I^{\otimes N-1} \otimes \vec {\sigma}, 
\end{eqnarray}
and for the partial and total spin operators,
\begin{equation}
\label{sms}
\vec {S}_{(\alpha)} \equiv \sum^{\alpha}_{\beta=A} \vec {S}_{\beta},~~ \alpha = A, B, \cdots, N;~~~\vec {S} \equiv \vec {S}_{(\alpha=N)}, 
\end{equation}
the following spin invariances hold \cite{VLPT}:
\begin{equation}
\label{comm}
\left[\vec {S}^2_{(\alpha)}, \rho^{\otimes N}\right]=0~~~~~~~ \alpha=A,\cdots,N,
\end{equation}
and since 
\begin{equation}
\label{directsum}
\left[\vec {S}^2_{(\alpha)}, \vec{S}^2_{(\beta)}\right]=0\qquad \forall \alpha,\beta 
\end{equation}
the total Hilbert space can be written as a direct sum
\begin{equation}
\label{totalh}
{\cal H}^{(N)}= \oplus_{\{ s_{(\alpha)}\} } E_{\{ s_{(\alpha)}\} } 
\end{equation}
where $E_{\{ s_{(\alpha)}\} }$ are the simultaneous eigenspaces of all the operators $\vec{S}^2_{(\alpha)}, \forall \alpha \not = A$, with corresponding eigenvalues $\{ s_{\alpha}(s_{\alpha} +1)\}$, ordered with decreasing $\alpha$ (see \cite{VLPT} for more details). For instance, for $N = 2$ only $\vec{S}^2_{(B)}$ ($s_{(B)}$) is relevant, i.e. $E_{\{ s_{(\alpha)}\} } = E_{s_{(B)}}$, and the decomposition reads
\be
{\cal H}^{(N=2)}= E_1 \oplus E_0,
\ee
where $E_1$ is the triplet or symmetric (under exchange of copies) subspace, with total spin $s\equiv s_{(B)}=1$, whereas $E_0$ is the singlet or antisymmetric subspace, with total spin $s=0$. Then,
\vspace{3mm}
 
\noindent{\bf Corollary 2:} There always exists an optimal measuring strategy consisting only of rank-one operators of the form $\proj{\{s_{(\alpha)}\}}$, where the not necessarily normalized vector $\ket{\{s_{(\alpha})\}}$ is an eigenvector of all partial and total spin operators, i.e.
\be
\vec{S}^2_{(\beta)}~ \ket{\{s_{(\alpha)}\}}~ =~ s_{(\beta)}(s_{(\beta)}+1)~  \ket{\{s_{(\alpha)}\}} ~~~\forall\beta,
\ee
and thus it belongs to the subspace $E_{\{s_{(\alpha)}\}}$.
\vspace{3mm}

\noindent{\bf Proof:} Let $\sum_i M_i=\openone$ correspond to an optimal measurement with rank-one operators $M_i = \proj{i}$ (where the \ket{i} do not need to be orthogonal nor normalized) and let $\Pi_{\{s_{\alpha}\}}=\Pi_{\{s_{\alpha}\}}^{2}$ be a projector onto the whole subspace $E_{\{s_{\alpha}\}}$. Then it follows from eq. (\ref{totalh}) that
\be
\sum_{\{s_{\alpha}\}} \Pi_{\{s_{\alpha}\}} \openone \Pi_{\{s_{\alpha}\}} = \sum_{\{s_{\alpha}\}} \Pi_{\{s_{\alpha}\}} = \openone,
\ee
so that if we replace $\openone$ with $\sum_i M_i$ in the LHS of this equation, we obtain a new measurement
\be
\sum_{i,\{s_{\alpha}\}}\proj{i,\{s_{\alpha}\}} = \openone;~~~~~ \ket{i,\{s_{\alpha}\}}\equiv \Pi_{\{s_{\alpha}\}} \ket{i}.
\label{optimtoo}
\ee
Now, since eq. (\ref{comm}) implies that for each $\ket{i}$,
\be
\mbox{Tr}[\rho(\vec{b})^{\otimes N}\proj{i}] = \sum_{\{s_{\alpha}\}} \mbox{Tr}[\rho(\vec{b})^{\otimes N}\proj{i,\{s_{\alpha}\}}],
\ee
the theorem guarantees that the measurement of eq. (\ref{optimtoo}) is also optimal. $\Box$

 [Notice that exactly the same conclusion was also achieved, for any $N$, when the fidelity was used as a criterion for optimality \cite{VLPT}, this being indicative of the universality we are considering here.]

 Thus, in order to find an optimal measuring strategy for $N=2$ we can always choose the pure states on which the measurement projects to be symmetric or antisymmetric under the exchange of the two qubits. Let us next compute $\Delta I^{(2)}$ for the optimal strategy of ref. \cite{VLPT}, that is corresponding to a resolution of the identity of the form
\be
\openone = \proj{\sigma} + \frac{3}{4}\sum_{i=1}^4(\proj{\hat{n}_i})^{\otimes 2},
\label{opti2}
\ee
where $\ket{\sigma}$ is the (normalized) singlet state, $\vec{\sigma}\cdot\hat{n} \ket{\hat{n}} = \ket{\hat{n}}$ ($\braket{\hat{n}}{\hat{n}}=1$) and the four unitary vectors $\hat{n}_i$ point to the four directions of the vertices of a regular tetrahedron. One readily obtains
\bea
f_c(\vec{b}|\sigma) = \frac{1-b^2}{4} \frac{f(b)}{P_{ap}(\sigma)}~&;&~P_{ap}(\sigma) = I_1\\
f_c(\vec{b}|\hat{n}) = \frac{3}{16} (1+\vec{b}\cdot\hat{n})^2 \frac{f(b)}{P_{ap}(\hat{n})}&;&P_{ap}(\hat{n}) = \frac{1}{4}(1-I_1)
\eea
so that
\bea
\Delta I^{(2)} = \pi\int_0^1db~b^2f(b)\left[ \frac{(1+b)^3}{b}\log_2(1+b) \right.-\nonumber \\
\left. \frac{(1-b)^3}{b}\log_2(1-b) + (1-b^2)\log_2(1-b^2)\right] -\nonumber\\
(1-I_1)(\frac{2\log_2 e}{3} + \log_2\frac{1-I_1}{3}) -I_1\log_2I_1 - 2.
\label{expressio}
\eea
Can we do better, i.e. is there another resolution of the identity which leads to a larger $\Delta I^{(2)}$? Let us prove that there is none. Because of corollary 2, the whole question boils down to whether symmetric entangled states could do better than the symmetric product states $\ket{\hat{n}_i}\ket{\hat{n}_i}$ used in eq. (\ref{opti2}). Consider therefore a general symmetric state of Schmidt decomposition
\be
\ket{\psi} =\sqrt{p} \ket{+}\ket{+} + \sqrt{1-p}\ket{-}\ket{-},~~~p\in[0,1],
\ee
where the isotropy of $f(b)$ has been taken into account in choosing the basis. One can readily obtain the average Kullback information corresponding to this state,
\[
\Delta I^{(2)}_{\psi} = \frac{1}{2}\int_0^1db~b^2f(b)\int_0^{2\pi}d\phi\int_{-1}^{1}d\mu ~h\log_2\frac{h}{\frac{1-I_1}{3}},
\]
\bea
h &\equiv& k + l\cos 2\phi,~~~~~l \equiv 2\sqrt{p(1-p)}b^2(1-\mu^2),\nonumber\\
~ k &\equiv& 1+b^2\mu^2 + (2p-1)2b\mu,
\eea
which after integration of $\phi$ gives
\bea
\Delta I^{(2)}_{\psi} &=& \frac{\pi}{2}\int_0^1db~b^2f(b)\int_{-1}^{1}d\mu \{(1+b^2\mu^2)\log_2 \frac{3e}{2(1-I_1)} \nonumber \\
&+& k\log_2 (k + \sqrt{k^2 - l^2})\}.
\eea
 This is a function of $p$ that we want to maximize. Only $k\log_2 (k + \sqrt{k^2 - l^2})$ depends on $p$. The part $- l^2$ is maximized for $p=0$ and $p=1$. The other part too, as one can see easily neglecting the term $l^2$. Thus $\Delta I^{(2)}_{\psi}$ is maximized when $\ket{\psi}$ is a product state and the resolution of eq. (\ref{opti2}) is indeed optimal.

As we did for $N=1$, it is interesting to recall, with the help of ref. \cite{VLPT}, the average increase in fidelity for $N=2$
\be
\Delta F^{(2)} = \sqrt{(I_{\frac{1}{2}}-I_{\frac{3}{2}})^2 + \frac{1}{16}(1-4I_1)^2} + I_{\frac{3}{2}} - I_{\frac{1}{2}}.
\ee
One can now check that both $\Delta I^{(2)}$ and $\Delta F^{(2)}$ are again maximized for the distribution eq. (\ref{pure}). For $\Delta I^{(2)}$ this follows by observing that the part in square brackets of eq. (\ref{expressio}) is an increasing function of $b$ and that the other part, which depends on $I_1$, increases as $I_1$ goes towards zero.

 We have thus checked for $N=1$ and $N=2$ that both the fidelity and the Kullback information lead to the same optimal measuring strategy and to the same, pure state, distribution which maximizes their increases. We conjecture, while not foreseeing any feature which could jeopardize extending the proof to $N>2$, that the universality of optimal measurements holds for any number $N$ of copies of the unknown state \cite{localmax}. Corollary 2 makes this conjecture very plausible. The precise optimal strategy is in fact determined to a great extend by the isotropy of the prior distribution, the symmetries of the state $\rho(\vec{b})^{\otimes N}$ which allow to choose each positive operator $M_i$ to act only on one of the subspaces $E_{\{s(\alpha)\}}$, and the fact that both the fidelity and the Kullback favour strategies with outcomes $i$ whose normalized probability of occurrence Tr$[\rho(\vec{b})^{N} M_i]/$Tr$[M_i]$ spans the largest possible range as a function of the direction of $\vec{b}$.

 Now, suppose we want to use the $N$ qubits as a quantum channel of classical information. Alice prepares $N$ copies of a given state $\rho(\vec{b})$ (the classical information being encoded in the vector $\vec{b}$) and sends them to Bob, who will perform a collective measurement in order to recover as much information about $\vec{b}$ as possible. The previous results single out using, when restricted to isotropic prior distributions, only pure states $(b=1)$ to encode classical information as the optimal method. We can then easily compute the optimal capacity of this isotropic quantum channel for any $N$, to find that
\be
\Delta I^{(N)} = \log_2 (N+1) - \frac{N}{(N+1)}\log_2 e,
\ee
which for large $N$ gives $\frac{\log_2 N}{N}$ bits carried per qubit. Notice that this is a purely quantum channel, no additional flow of classical information being required at any stage. Its poor capacity can be exponentially enhanced without spoiling this fact if we take into account that a pure state $\phi^{\otimes N}$ belongs to the symmetric subspace ${\cal S}^{(N)}$ of the whole Hilbert space ${\cal H}^{(N)}$. Since the dimension of ${\cal S}^{(N)}$ is $N+1$, which corresponds to the dimension of a Hilbert space ${\cal H}^{(M)}$ of $M \equiv \log_2 (N+1)$ qubits, Alice can always compress, by means of a state-independent, unitary (and thus fully reversible) transformation, the state $\phi^{\otimes N}$ to fit in $M$ qubits, that will then be transferred to Bob. In this case the capacity increases up to $1 - O(\frac{1}{\log N})$ bits per qubit, which is asymptotically the classical one (as expectable, since for any two inequivalent states $\phi$ and $\phi'$, $\phi^{\otimes N}$ and $\phi'^{\otimes N}$ become orthogonal as $N \rightarrow \infty$), and which is consistent with the Levitin-Holevo bound \cite{Holevo} for the classical capacity of a quantum channel.
  
 Summarizing, using the gain of information as a guide we have constructed optimal and minimal measurements on $N = 1,2$ identical copies and have shown that for isotropic distributions the maximal gain of information is achieved for pure states. Also universality of optimal measurements has been proven, since these measurements exactly coincide with those obtained in previous work, where the fidelity was taken as figure of merit. We conjecture that also for $N \geq 3$ the most informative measurements are the most faithful ones, and vice versa.

 G.V. acknowledges a CIRIT grant 1997FI-00068 PG. Financial support from CIRYT, contract AEN98-0431, CIRIT, contract 1998SGR-00026 and from the ESF-QIT programme is also acknowledged.


\begin{references}
\bibitem{text} Furthermore, and due to the symmetry in the $i$ and $\vec{b}$ distributions of Bayes' formula, eq. (\ref{Bayes}), the expressions in eqs. (\ref{avekullback}) and (\ref{deltaentro}) are also equal to the corresponding expressions for which $f$ and $f_c$ have been traded for $P_{ap}$ and $P$.
\bibitem{localmax} We have also been able to check, for an arbitrary number $N$ of copies of the unknown state, that the optimal measurements according to the fidelity (as presented in \cite{VLPT}) are at least locally optimal (that is, better than any other measurement which follows from infinitessimally perturbing the former) for the Kullback.
\end{references}
\end{document}